# The Millennial Boom, the Baby Bust, and the Housing Market[*]

**Marijn A. Bolhuis**

University of Toronto

marijn.bolhuis@mail.utoronto.ca

**Judd N.L. Cramer**

Harvard University

cramer@fas.harvard.edu

March 22, 2020

**Abstract**

As baby boomers have begun to downsize and retire, their preferences now overlap with millennial's predilection for urban amenities and smaller living spaces. This confluence in tastes between the two largest age segments of the U.S. population has meaningfully changed the evolution of home prices in the United States. Utilizing a Bartik shift-share instrument for demography-driven demand shocks, we show that from 2000 to 2018 (i) the price growth of four- and five-bedroom houses has lagged the prices of one- and two-bedroom homes, (ii) within local labor markets, the relative home prices in baby boomer-rich zip codes have declined compared with millennial-rich neighborhoods, and (iii) the zip codes with the largest relative share of smaller homes have grown fastest. These patterns have become more pronounced during the latest economic cycle. We show that the effects are concentrated in areas where housing supply is most inelastic. If this pattern in the housing market persists or expands, the approximately $16.5 trillion in real estate wealth held by households headed by those aged 55 or older will be significantly affected. We find little evidence that these upcoming changes have been incorporated into current prices.

[*] First version: February 10, 2020. We thank Larry Summers for extensive feedback and valuable discussions. Bolhuis gratefully acknowledges financial support from the Ontario Trillium Foundation.

## I. Introduction

That the emergence and recession of baby boomers as drivers of the American population would have large effects on the housing market has been hypothesized since at least Mankiw and Weil (1989, 1992). While these early predictions of nation-wide housing price declines as baby boomers entered middle age in the 1990s were erroneous, this brief uses granular, neighborhood/housing structure data scraped from Zillow to track the emergence of broad generational patterns in the U.S. housing market. As baby boomers have begun to downsize and retire, their preferences now overlap with millennial's predilection for urban amenities and smaller living spaces. Utilizing a Bartik shift-share instrument for demography-driven demand shocks, we show that from 2000 to 2018

- (i) the price growth of four- and five-bedroom houses has lagged the prices of one- and two-bedroom homes,
- (ii) within local labor markets, the relative home prices in baby boomer-rich zip-codes have declined compared with millennial-rich neighborhoods, and
- (iii) the zip codes with the largest relative share of smaller homes have experienced the fastest price growth.

These patterns have become more pronounced during this economic cycle (2012-2018). We show that the effects are concentrated in areas where housing supply is most inelastic. Price simulations using demographic projections from 2000 fit the period through 2016 well. If this pattern in the housing market persists or expands, the approximately $16.5 trillion in real estate wealth held by households headed by those aged 55 or older will be significantly affected. We find little strong evidence that these upcoming changes have been incorporated into current prices.

The economic literature has not kept pace with the advancement of data on housing characteristics at the neighborhood level, nor has it considered the recent transition into retirement age of the baby boomers. Romem (2019) discusses the "silver tsunami" of boomers leaving their homes in the future and outlines which metropolitan areas would be most affected by this development, but this research does not engage the role of downsizing and the relatively smaller household sizes of millennials. Other studies focus on the effect of demographic change on housing preferences (Couture & Handbury, 2017), on geographic variation in price-rent ratios (Begley et al., 2019) or on aggregate home prices in other countries (e.g. Hiller & Lerbs, 2016), but do not consider its effect on different segments of the housing market.

The rest of this paper proceeds as follows. Section II provides a background and summarizes the data. Section III provides the empirical findings. Section IV concludes.

## II. Background and Data

**Demographic Trends and the Housing Market**

The US has been, and still is, undergoing demographic change that impacts the demand side of the housing market. Our Figure 1 below, constructed using data from the American Community Survey (ACS), shows the rising share of new homes purchased by those 60 and older. Note how the share of smaller homes purchased by this group has risen relatively faster. Their share of one- and two-bedrooms has almost doubled in the last twenty years.

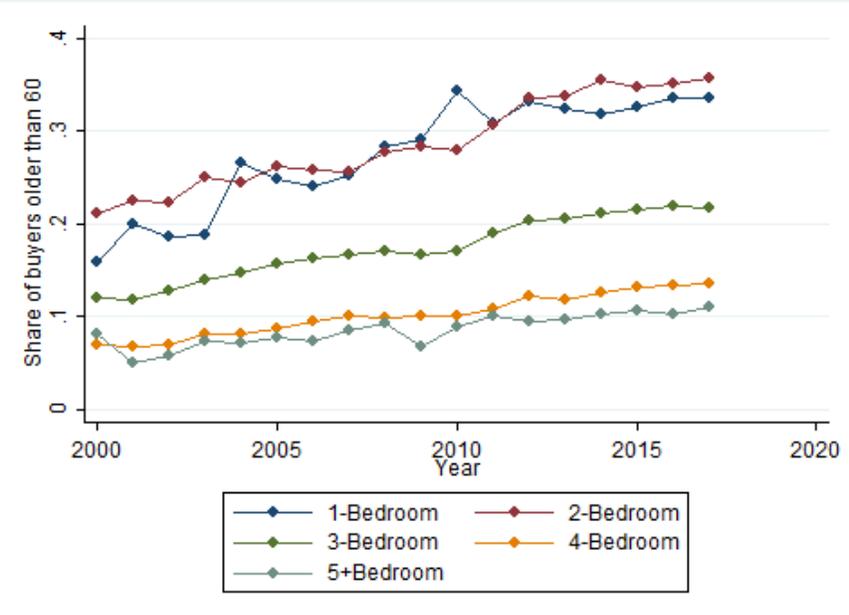

**Figure 1:** plots share of homes bought by those aged 60 and older, for different home types. Data from ACS, 2000-2017.

As buyers age, relative demand has shifted towards smaller homes. Over the life cycle, the demand for housing tends to peak around age 40, and decline thereafter (Figure 2). The relative demand for one- and two-bedrooms has therefore increased substantially in the last twenty years, driven by the higher share of millennials and baby boomers in the US population (Figure 3). We expect these trends to continue. For example, the adult population share of 70+-year-olds is expected to grow at an even faster rate in the coming decade, increasing from 15 to 20 percent of the U.S. population in the next ten years (Appendix).

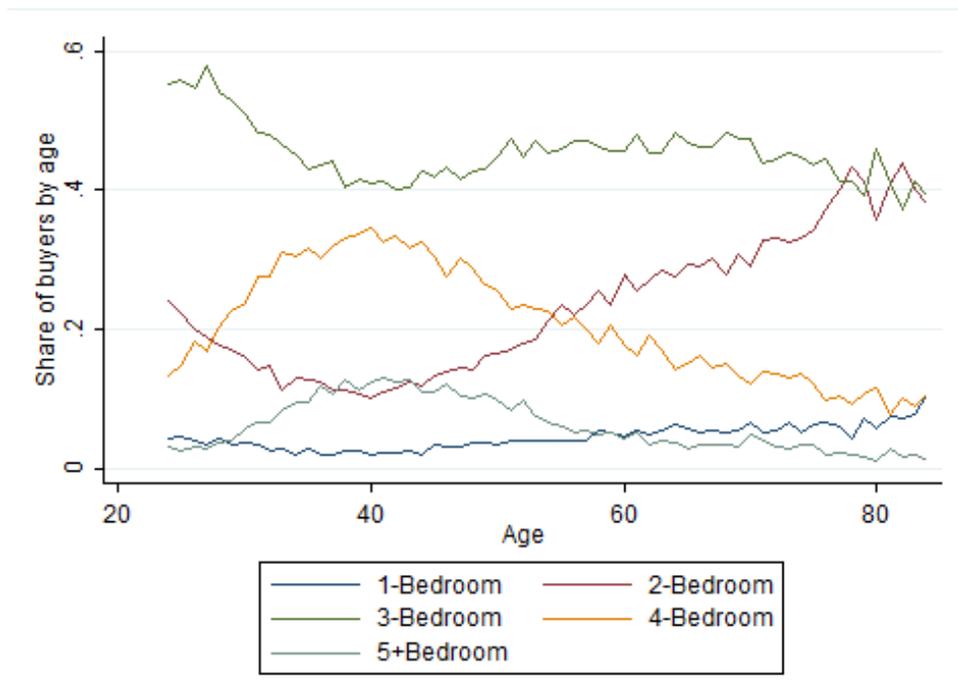

**Figure 2:** Life-cycle demand for home types. Data from ACS, 2017.

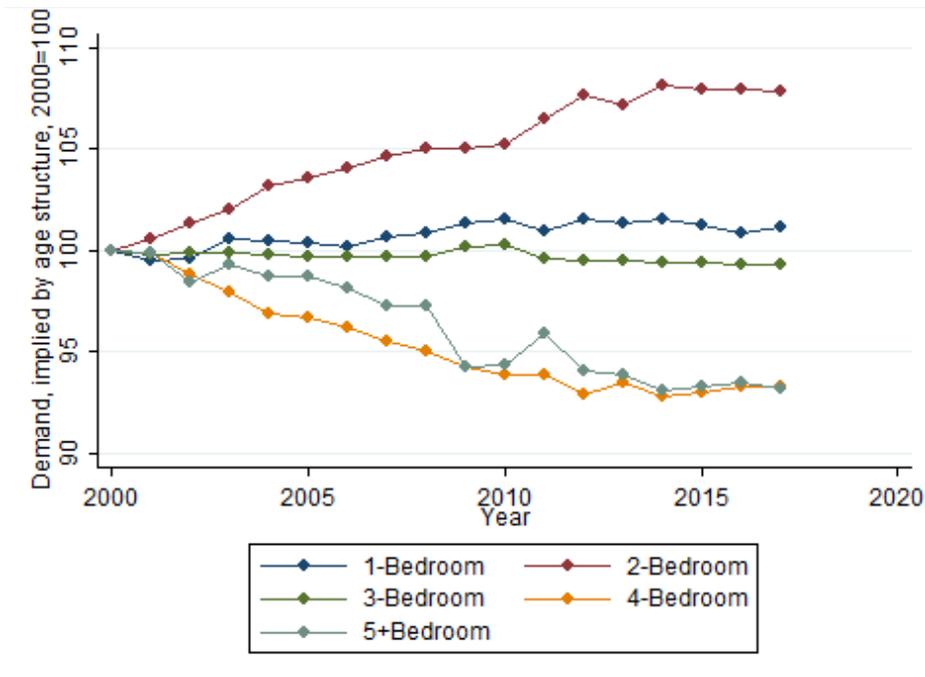

**Figure 3:** Changes in relative demand for different home types, as implied by the census age composition and home preferences from 2000. Data from ACS, 2000-2017.

**Local Housing Market Data and Empirical Strategy**

We combine detailed zip code-level housing market data with county-level statistics on demographic changes to test the effect of demographic demand shock on neighborhood's home prices, turnover and rents. We start by scraping public data from Zillow on house prices for different types of houses (1-bedroom, detached, multifamily, and so on) at the zip code level. In turn, these data are combined with geo-coded census data tables from IPUMS NHGIS, both at the county and zip code level. The final dataset that we use in the empirical analysis contains 14,653 unique zip codes, covering almost 80% of American owned homes in 2000.

The changes in relative housing demand match the diverging trends in prices of different types of homes over the current cycle. Within neighborhoods, price growth of one-bedrooms has outpaced that of 5-bedrooms by almost two percentage points per year (Figure 4a). Across neighborhoods, annual price growth in zip codes with less than 20% of their housing stock in

4- and 5+bedrooms has been almost two percentage points higher than in areas with more than 30% of their stock in this segment (Figure 4b).

We combine our data to construct a demographic demand shift-share Bartik instrument. We do this by merging census data on county-level population changes by age group and zip code level home type distribution. Our Bartik uses a neighborhood's composition of home types as shares and county-level changes in the age composition as shifts. As such, we construct zip code-level within-county changes in demand for housing that are driven by the initial housing distribution.[†]

We now outline the regression specifications and details on the construction of the Bartik instrument. First, we construct the shock to housing demand in zip code $z$ located in county $c$ and state $s$. Let $d^b_{cs,t,t-1}$ be the percentage increase in population in county $c$ that chooses to live in housing type $b$ between period $t$ and $t$-$1$:

$$d^b_{cs,t,t-1} = 100 * \left( \frac{\sum_{a=1}^{A} s^{b,a} * P^a_{cs,t}}{\sum_{a=1}^{A} s^{b,a} * P^a_{cs,t}} - 1 \right):$$

where $s^{b,a}$ is the share of age group $a$ that chooses to live in housing type $b$ (using 14 census age groups, 20-85+, from the 2000 ACS), and $P^a_{cs,t}$ is the population of age group $a$ in county $c$ at time $t$. We interact this county-level population shifter with the zip code-level stock of housing to construct the zip code-level demand shock $D_{zcs,t,t-1}$:

$$D_{zcs,t,t-1} = \sum_{b=1}^{B} s^b_{zcs} * d^b_{cs,t,t-1}$$

where $s^b_{zcs}$ is the share of homes in zip code $z$ that is of housing type $b$, where the housing type denotes the number of bedrooms (1, 2, 3, 4, or more than 5).

In our regression framework, we run:

---

[†] This Bartik is analogous to a labor demand shift-share instrument (e.g. Autor et al., 2013) that uses within-region (e.g. local labor markets) industry composition as shares and regional (e.g. state) industry growth rates as shifts.

$$y_{zcs,t,t-1} = \gamma_{cs,t} + \beta * D_{zcs,t,t-1} + \epsilon_{zcs,t,t-1}$$

where $y_{zcs,t,t-1}$ is the percentage change in the dependent variable -often the home price index- in zip code $z$ in country $c$ in state $s$ between time $t$ and $t-1$. $\gamma_{zcs,t}$ is a county FE capturing common local labor demand shocks, legal restrictions, financing conditions, changing amenities, etc. $D_{zcs,t,t-1}$ is the Bartik housing demand shock defined above, and $β$ is the elasticity of interest. $\epsilon_{zcs,t,t-1}$ is a zip code-specific error term.

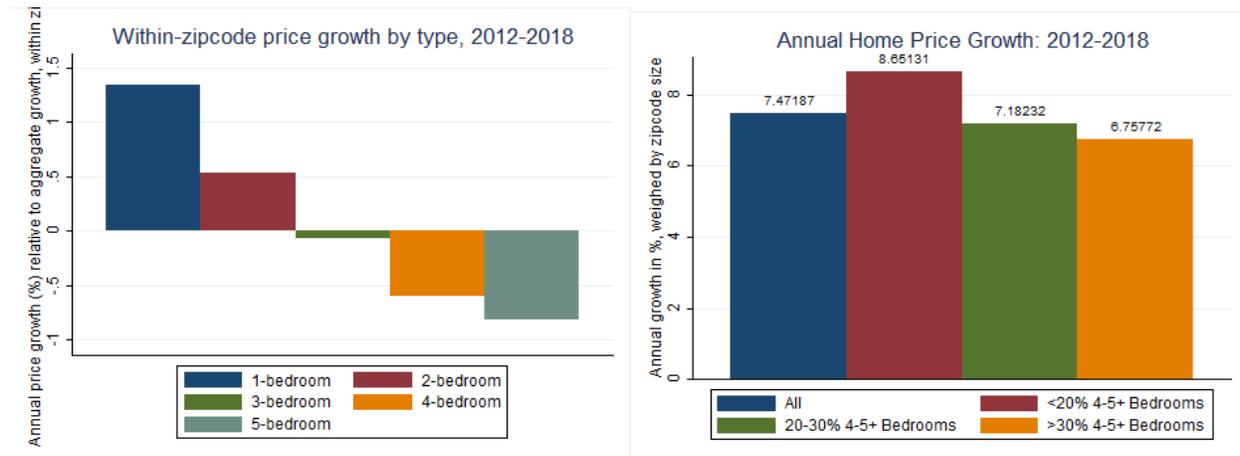

**Figure 4:** Price growth within and between neighborhoods, 2012-18. Data from Zillow.

# III. Empirical Findings

## Main Results

We document a robust correlation between changes in housing demand induced by demographic developments and zip code-level home price growth. While it is clear that there are large shifts underway, it remains challenging to distinguish between changes in demographics, preferences (Couture & Handbury, 2017), or a combination of the two as underlying causes. We err on the side of demographics as the patterns we report are completely consistent with a demographic demand shock, whereas there is little evidence in the census data on a broad preference shift towards smaller homes.

We can demonstrate that even in cities that attract young people, price pressure is highest in areas where relatively fewer boomers live and that contain homes with relatively fewer bedrooms. Even within the same county, the neighborhoods (zip codes) with relatively fewer 4+ bedroom houses have experienced faster price growth. The figure below presents the 6 biggest counties by population; each dot is 5 binned zip codes. In these 6 largest counties, prices in zip codes with the lowest share of 4+ bedrooms have grown ~1.5 as fast as prices in zip codes with the highest share of 4+ bedrooms.

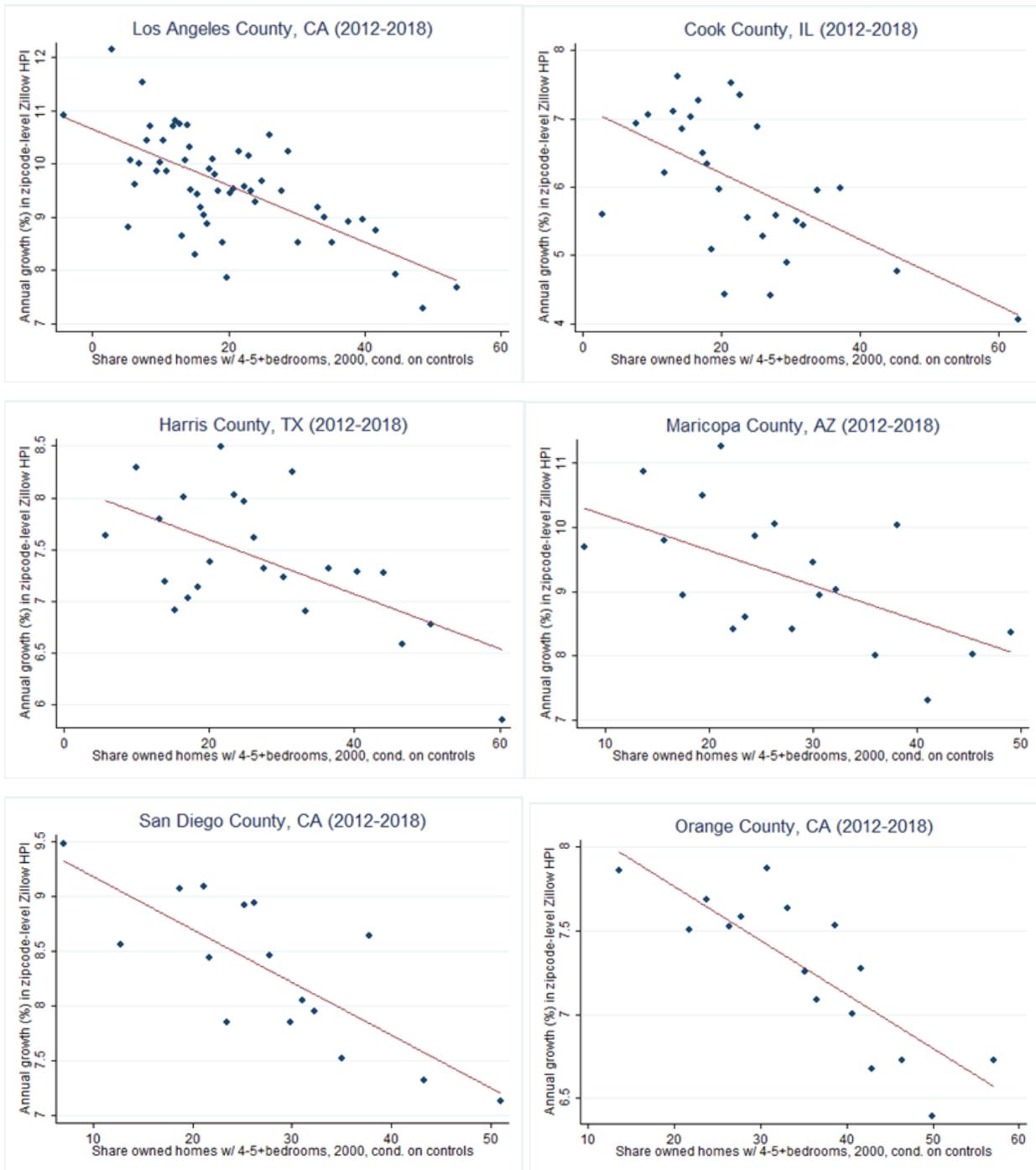

**Figure 5:** Annual growth in Zillow's Home Price Index (2012-18) against share of owned homes with 4+-bedrooms, conditional on zip code-level controls and county fixed effects.

Our Bartik analysis estimates a large, positive and significant elasticity of zip code-level home price growth with respect to demographic demand shocks. Table 1 plots the corresponding regressions. The estimated elasticity is not very large from 2000-06, which is unsurprising in light of the fact that most boomers were not of retirement age yet during this time. The effect is quite strong, however, in the post-crisis period (2012-18), when boomers began to retire, and in the entire period (2000-18) overall. These effects are somewhat smaller for the subsample of counties with below median density and incomes, but even in those, the effect remains in line with our expectations (Table 2).

|  | (1) 2000-2006 | (2) 2000-2006 | (3) 2012-2018 | (4) 2012-2018 | (5) 2000-2018 | (6) 2000-2018 |
|---|---|---|---|---|---|---|
| Annual growth (%) zip code-level demand | 0.857 | 0.827 | 6.251*** | 5.675*** | 1.162*** | 1.325*** |
|  | (1.983) | (1.765) | (0.646) | (0.627) | (0.362) | (0.345) |
| Observations | 14,653 | 14,653 | 14,653 | 14,653 | 14,653 | 14,653 |
| R-squared | 0.756 | 0.762 | 0.709 | 0.713 | 0.668 | 0.693 |
| County FE | YES | YES | YES | YES | YES | YES |
| Controls | NO | YES | NO | YES | NO | YES |

**Table 1:** Dependent variable is annual growth (%) in zip code-level Zillow Home Price Index. Standard errors clustered at county level. Controls include zip code-level changes in share of residents that: identify as white; identify as black; commute using own vehicle; commute less than 20 minutes; have a college degree; are a student; are below the poverty line; are employed; work in construction; work in manufacturing; work in government; live in a detached home; live in an attached one-family home. Controls also include zip code-level changes in (log) median household income, per capita income, as well as median rents.

| VARIABLES | (1) <p50 density | (2) <p50 density | (3) <p50 income | (4) <p50 income |
|---|---|---|---|---|
| Annual growth (%) zipcode-level demand | 1.847*** | 6.975*** | 3.773*** | 6.134*** |
|  | (0.528) | (0.878) | (0.783) | (0.810) |
| Observations | 6,924 | 7,389 | 6,937 | 7,384 |
| R-squared | 0.828 | 0.743 | 0.788 | 0.770 |
| County FE | YES | YES | YES | YES |
| Controls | YES | YES | YES | YES |

**Table 2:** Dependent variable is annual growth (%) in zip code-level Zillow Home Price Index. Standard errors clustered at county level. Controls include zip code-level changes in share of residents that: identify as white; identify as black; commute using own vehicle; commute less than 20 minutes; have a college degree; are a student; are below the poverty line; are employed; work in construction; work in manufacturing; work in government; live in a detached home; live in an attached one-family home. Controls also include zip code-level changes in (log) median household income, per capita income, as well as median rents.

The effects we find are mostly driven by zip codes with more/fewer millennials/baby boomers (Table 3). Given that millennials tend to live in one- or two-bedroom homes while boomers reside in four- and five-bedroom homes, this is not surprising. These differences are stark: neighborhoods with 20 percentage points more baby boomers experienced 0.1 percentage point lower annual price growth. For a visual overview, see the appendix, in which we also document that neighborhoods with more/fewer 1/2-bedrooms and 4+bedrooms drive our results.

| VARIABLES | 2012-2018 | 2012-2018 | 2012-2018 |
|---|---|---|---|
| Share (%) of population 55-74 ("Boomers") | -0.0481*** (0.00599) | | |
| Share (%) of population 35-54 ("Gen X") | | -0.000559 (0.00508) | |
| Share (%) of population 20-34 ("Millennials") | | | 0.0448*** (0.00470) |
| Observations | 14,337 | 14,337 | 14,337 |
| R-squared | 0.787 | 0.781 | 0.790 |
| County FE | YES | YES | YES |
| Controls | YES | YES | YES |

**Table 3:** Dependent variable is annual growth (%) in zip code-level Zillow Home Price Index. Standard errors clustered at county level. Controls include zip code-level changes in share of residents that: identify as white; identify as black; commute using own vehicle; commute less than 20 minutes; have a college degree; are a student; are below the poverty line; are employed; work in construction; work in manufacturing; work in government; live in a detached home; live in an attached one-family home. Controls also include zip code-level changes in (log) median household income, per capita income, as well as median rents.

**Heterogeneity**

It seems that the effects we find are driven by areas with lower supply elasticities. We see a larger difference in price growth (2012-18) between 2-bedrooms and 4+bedrooms in metro areas with a lower supply elasticity (estimates from Saiz, QJE2008). In Figure 6 below we estimate that 2-bedrooms grew ~2 percentage points faster than 4+bedrooms in the most rigid MSAs, and ~0 in the most flexible ones. Comparing different types of neighborhoods, we see that the predictive power (for home price growth) of a neighborhood's share of baby boomers (Figure 7) or 4+bedrooms (Figure 8) is significantly higher/lower in MSAs with below/above median supply elasticities. In effect, almost all predictive power of neighborhoods' share of baby boomers is accounted for by the 50% of MSAs with the lowest supply elasticities. These results support our conjecture that the stark differences in price growth across neighborhoods and home types are driven by relative demand shocks to homes with fewer bedrooms.

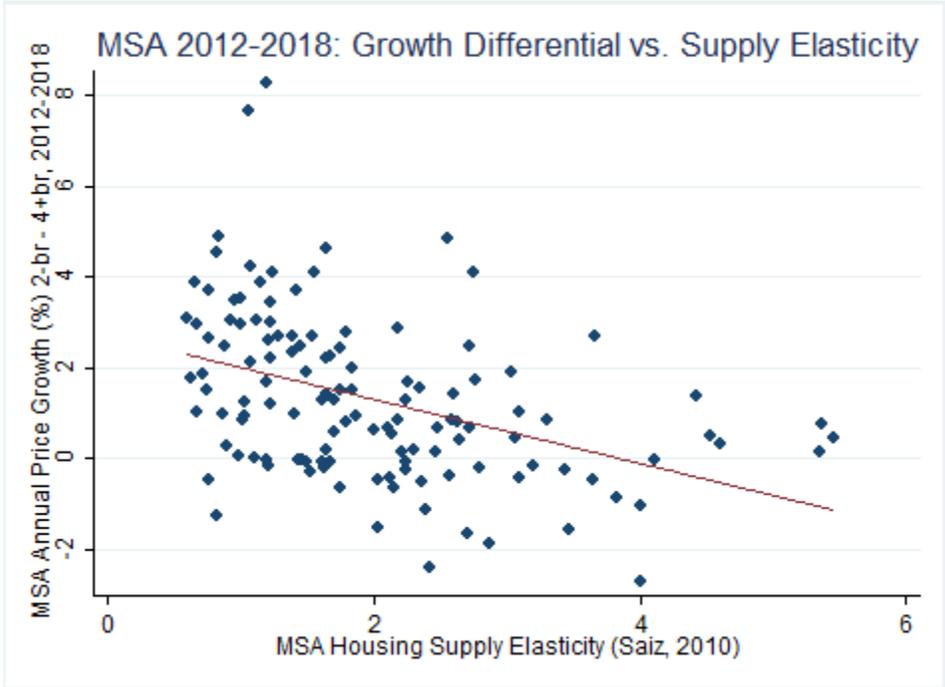

**Figure 6:** MSA-level difference in annual price growth of two- and four+ bedrooms (2012-18) against MSA housing supply elasticity of Saiz (2010).

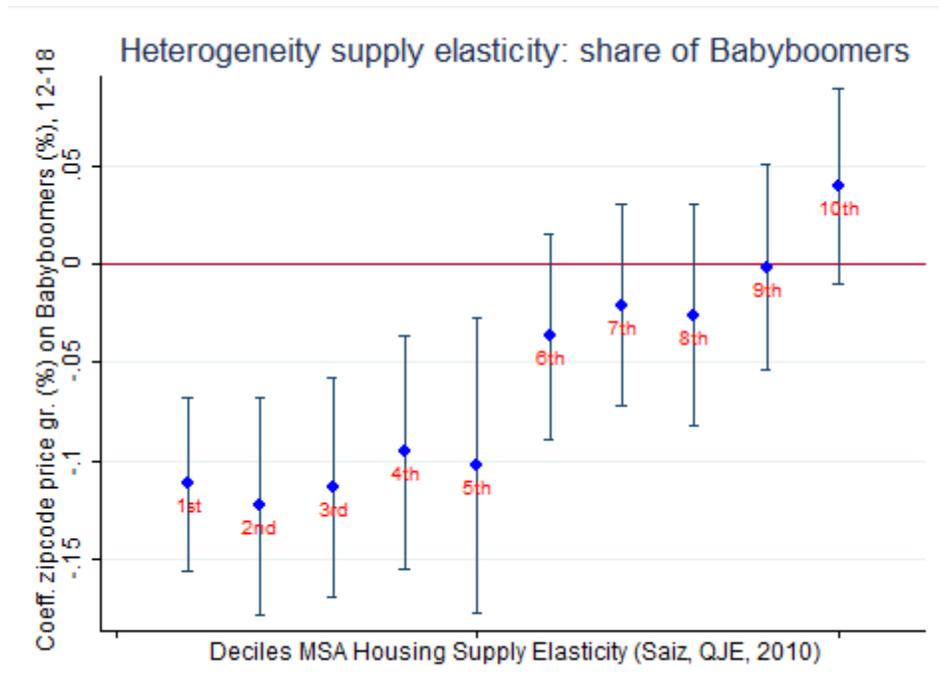

**Figure 7:** Heterogeneity in supply elasticity and neighborhood's share of baby boomers. Figure plots coefficient of regression with neighborhood price growth on its share of baby boomers.

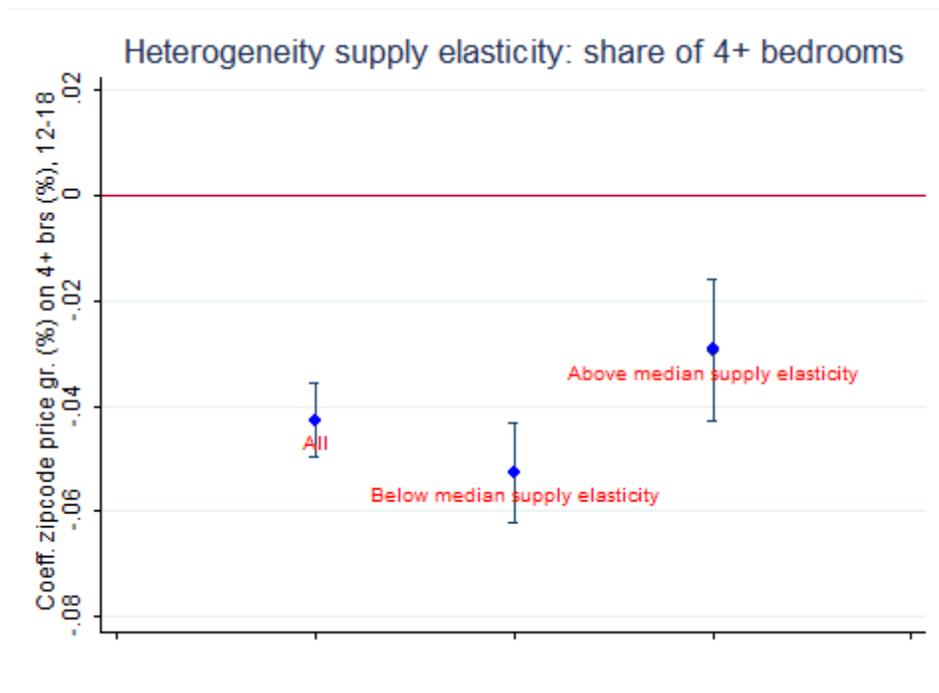

**Figure 8:** Heterogeneity in supply elasticity vs. neighborhood's share of 4+ bedrooms. Figure plots coefficient of regression with neighborhood price growth on its share of 4+ bedrooms.

The patterns we document are present in nearly all large US states. Figure 9 shows that the estimated elasticity of zip code-level home price growth with respect to the demand shocks is positive for all states with more than 150 neighborhoods, except Montana and Wisconsin.

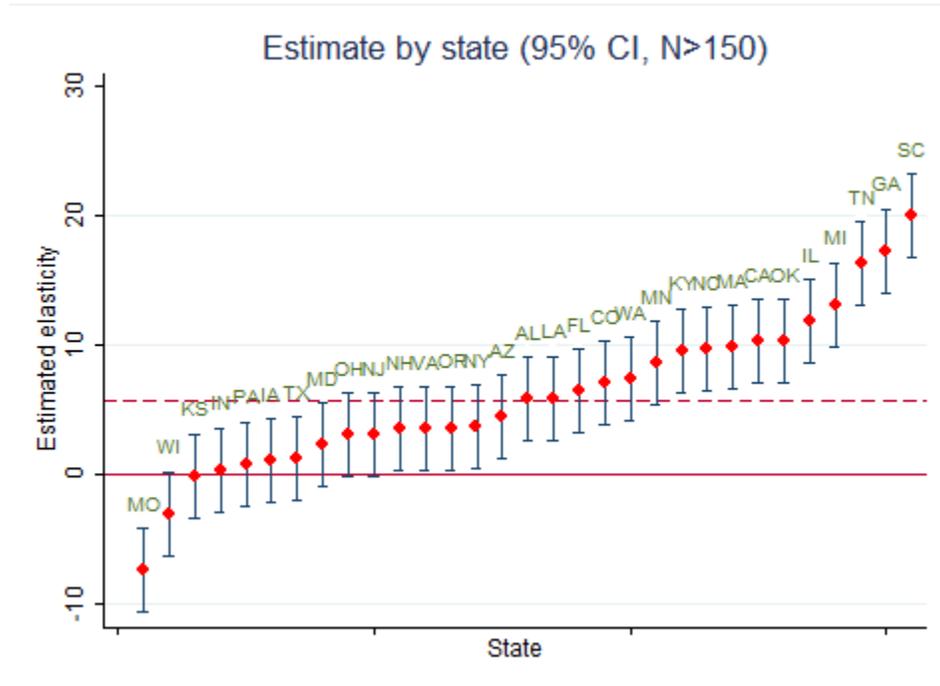

**Figure 9:** Heterogeneity by state, 2018-12. Figure plots estimated elasticity of main Bartik regression by states with more than 150 neighborhoods.

**Turnover, Selling Pressures and Expectations of Future Prices**

Turnover and liquidity have both increased significantly in neighborhoods that have experienced a positive demand shock relative to other areas. In Table 4 we regress annual growth in homes sold, for sale, newly listed and expected selling time on the zip code-level demand shock for 2012-18. While demand shocks are positively correlated with growth in new listings, they are negatively correlated with growth in *total* listings. As a result, the expected time to sell has decreased substantially in neighborhoods (e.g. with a high share of baby boomers) that experienced negative demand shocks. Moreover, Table 5 show that these houses in these neighborhood's are also more likely to be sold with a price cut, and that these price cuts tend to be larger. Moreover, these zip codes have shown slower growth in Zillow's

Buyer-Seller Index, which measures how 'hot' a market is relative to other regions and its own historical average.

|  | (1) Sold | (2) For sale | (3) For sale new | (4) Exp. sell time |
|---|---|---|---|---|
| Annual growth (%) zipcode-level demand | 11.14*** | -7.693*** | 26.17*** | -28.14*** |
|  | (1.659) | (1.368) | (8.356) | (6.240) |
| Observations | 7,909 | 7,909 | 310 | 310 |
| R-squared | 0.476 | 0.564 | 0.625 | 0.644 |
| County FE | YES | YES | YES | YES |
| Controls | YES | YES | NO | NO |

**Table 4:** Table presents results of regressions with annual growth (%) for 2012-2018 in monthly homes sold, homes for sale, new homes listed, and expected selling time. Usual controls apply.

|  | (1) Sold with cut (%) | (2) Med. price cut | (3) BSI region | (4) BSI time |
|---|---|---|---|---|
| Annual growth (%) zipcode-level demand | -10.25*** | -10.25*** | 36.27*** | 22.67*** |
|  | (1.879) | (1.879) | (4.773) | (4.570) |
| Observations | 7,221 | 7,221 | 7,221 | 7,221 |
| R-squared | 0.500 | 0.500 | 0.179 | 0.348 |
| County FE | YES | YES | YES | YES |
| Controls | YES | YES | YES | YES |

**Table 5:** Table presents results of regressions with annual growth (%) for 2012-2018 in percentage of homes sold with a price cut, median price cut, and the Zillow Buyer-Seller Index (BSI). The latter measure how hot a neighborhood's housing market is relative to other areas in the same region, and relative to its own history. The BSI uses three inputs: (i) percentage of listings with a price cut; (ii) median days on Zillow, and (iii) median sales-to-list price ratio. Usual controls apply.

These additional results suggest that owners in neighborhoods that experience slow demand growth have been anticipating further relative price decreases and are trying to get out of the market. We corroborate this conjecture by examining rent and price-rent growth in different neighborhoods in Table 6. Rents have increased significantly more in neighborhoods with higher demand growth, but this difference is small relative to that observed for home prices. As a result, price-to-rent ratios have diverged substantially between neighborhoods that have experienced differing demographic pressures. This might suggest sellers/buyers in these neighborhoods have been anticipating population-driven price changes (Begley et al., 2019). However, we should treat these data with some caution as the rent index is likely to reflect mostly one- and two-bedroom houses whereas the price index reflects the segment of the market with larger homes.

|  | (1) Price growth | (2) Rent growth | (3) Price-to-rent growth |
|---|---|---|---|
| Annual growth (%) zipcode-level demand | 7.157*** | 0.715* | 6.391*** |
|  | (0.829) | (0.370) | (0.872) |
| Observations | 7,316 | 7,316 | 7,316 |
| R-squared | 0.790 | 0.784 | 0.655 |
| County FE | YES | YES | YES |
| Controls | YES | YES | YES |

**Table 6:** Table presents results of Bartik regressions with annual growth (%) for 2012-2018 in home prices, rents, and price-to-rent ratios. Usual controls apply.

# IV. Conclusions

We have found evidence that suggests not only that large housing market corrections are underway, but also that these will be forthcoming in the next decade. The implications of these shifts are potentially big. Currently, Americans over 55 hold almost 60% of total housing wealth. For this group, 60-75% of their assets are in real estate (Davis & Van Nieuwerburgh, 2015). These aggregate statistics mask substantial heterogeneity: the American middle class is almost completely dependent on housing, with retirement portfolios as distant second. As a result, the asset position of the middle class tends to move closely with home prices (Kuhn et al., 2017). If the trends we document continue, a large share of the US population is at risk of taking a substantial hit to their savings while they move into retirement.

# Appendix

This Appendix contains additional empirical results.

- **Housing in the U.S. mainly differs in terms of the size of the unit (proxied by the number of bedrooms).** The overwhelming majority the owned stock of housing is detached (~80% of homes in 2017, ACS), whereas there is considerable heterogeneity in the number of bedrooms.

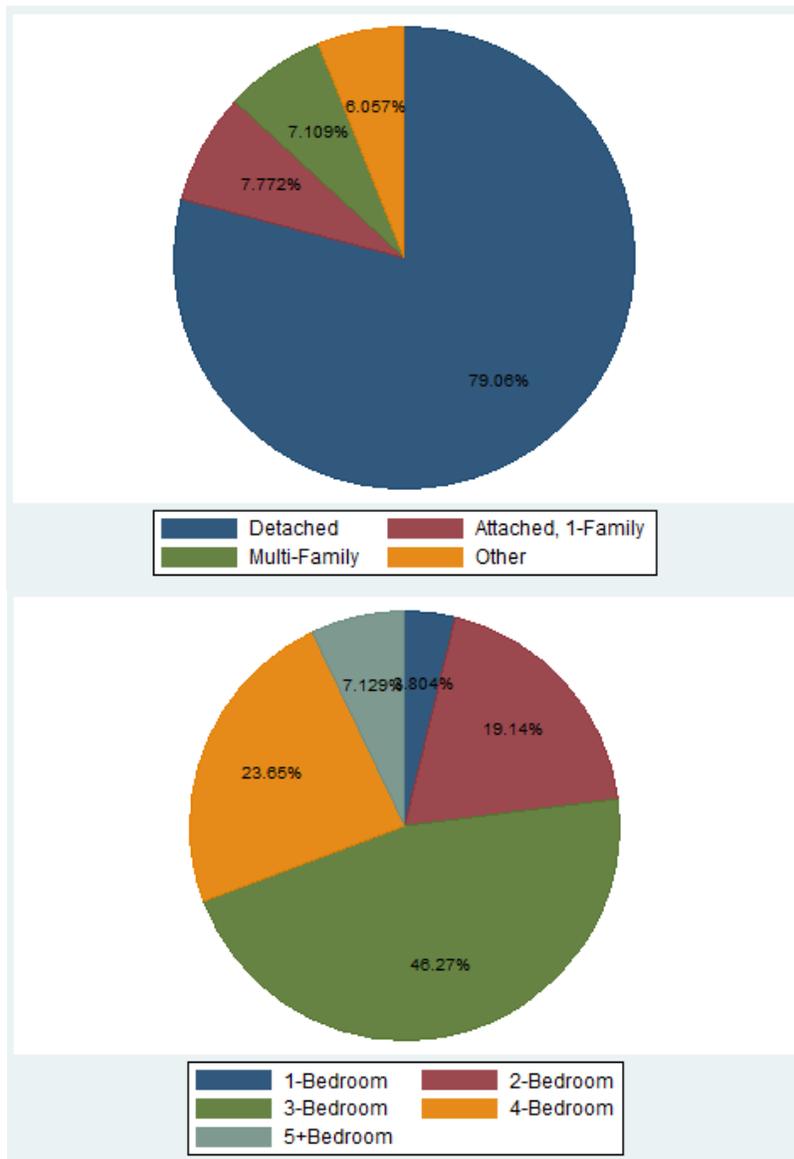

- **U.S. home ownership rates are monotonically increasing over the life cycle until owners reach 80, whereas buying activity in the housing market peaks around 30, and declines thereafter (ACS, 2017).**

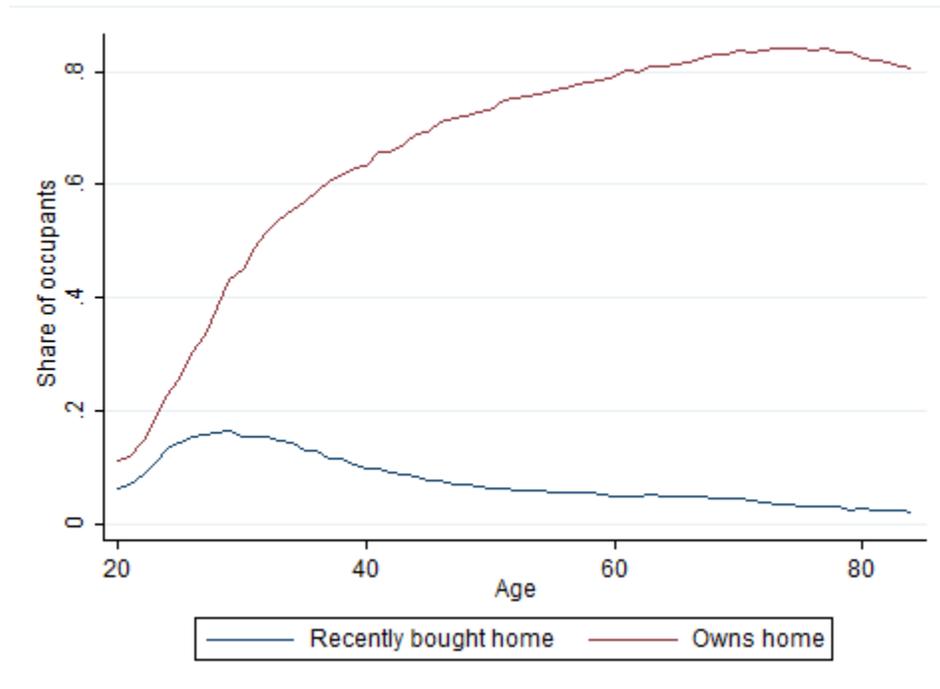

- **The U.S. housing has recovered from the bust in late 2000s and, averaging 4-4.5 % price growth per year between 2000 and 2018 (Zillow).**

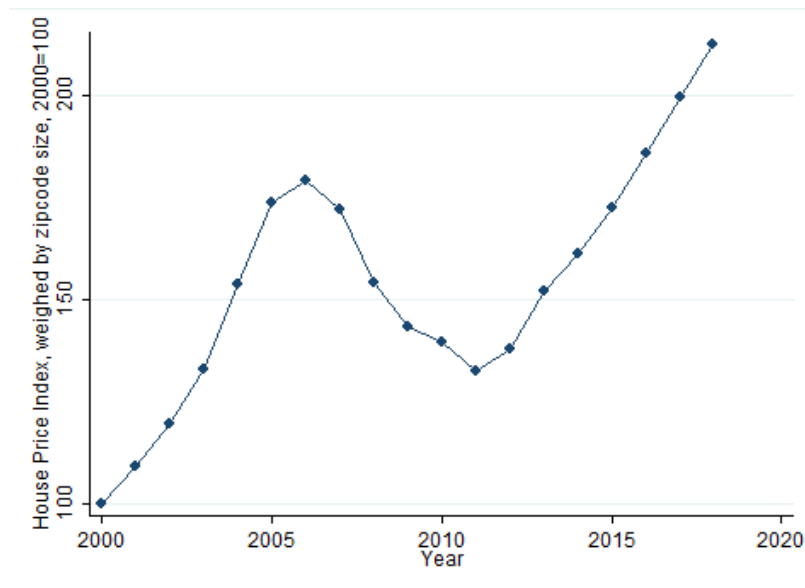

- **Over the same period, the mean age of recent buyers increased from about 42.5 to 46 (ACS), whereas the mean number of bedrooms of recently bought homes increased from 2.9 to 3.1 .**

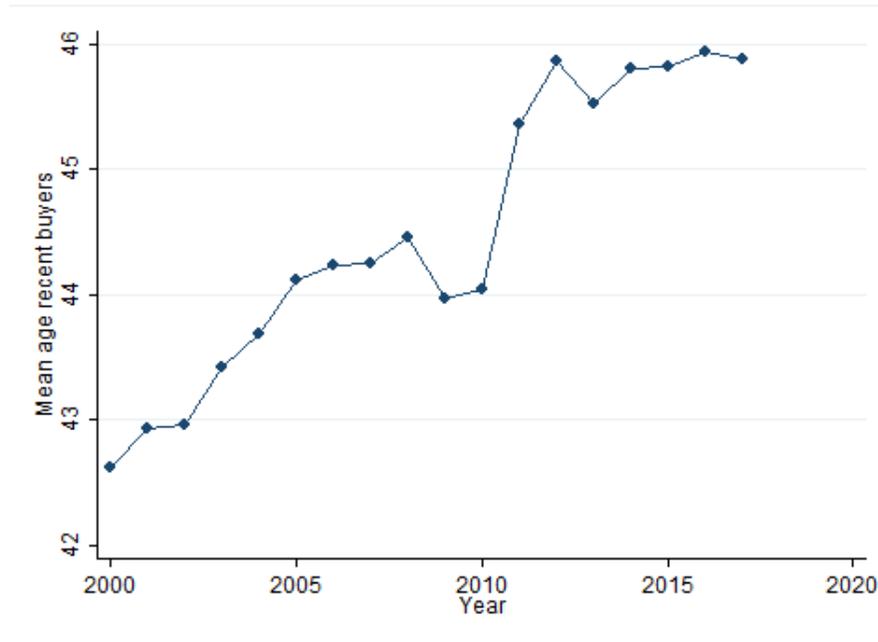

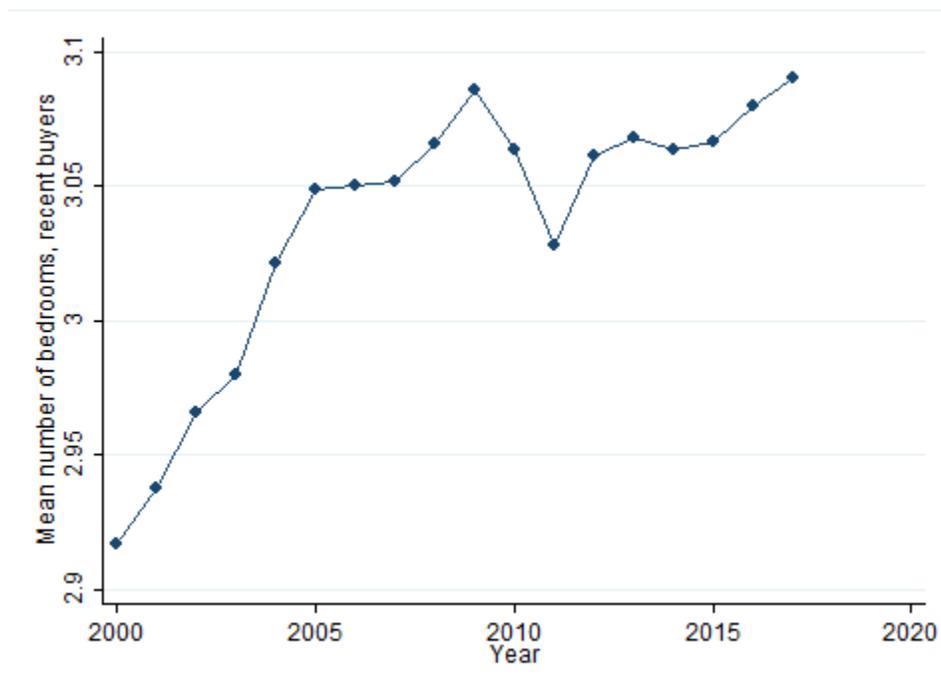

- **Based on demographic projections, we expect the relative demand for one- and two-bedrooms to keep increasing.** For example, the adult population share of 70+-year-olds is expected to grow at an even faster rate in the coming decade, increasing from 15 to 20 percent of the U.S. population in the next ten years.

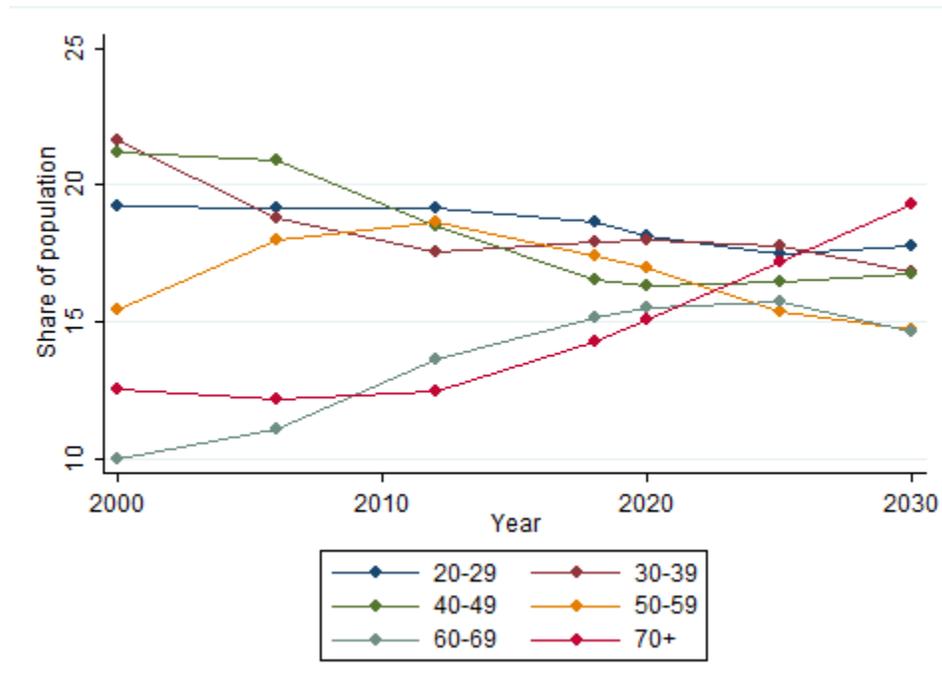

- **The lower price growth we find in zip codes with a higher share of 4+ bedrooms extends to all types of homes**. In the figures below, we plot home price growth for different types of homes against a neighborhood's share of homes with more than 4 bedrooms in 2000. These regressions used for these plots include county fixed effects, so these are within-county differences:

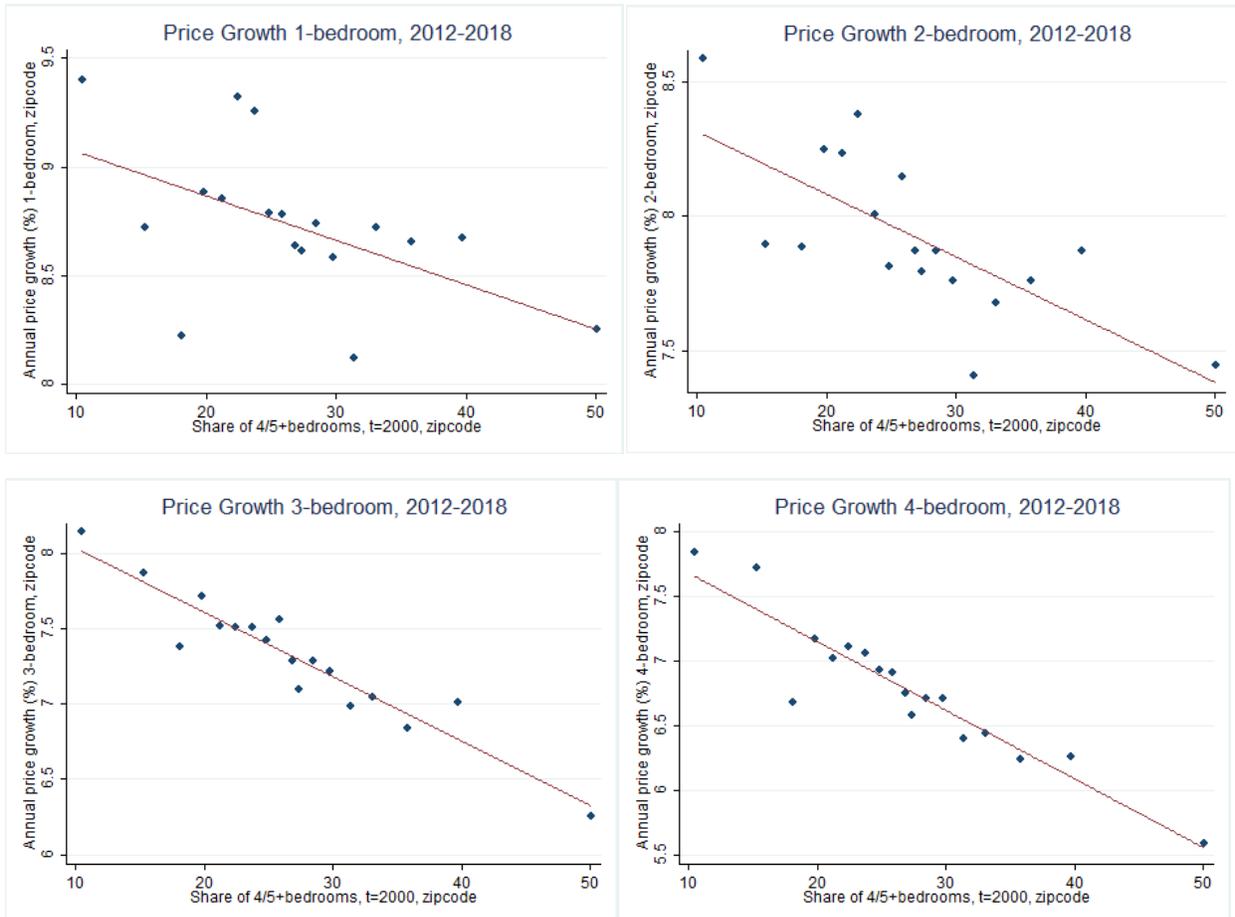